# Fabrication and properties of luminescence polymer composites with erbium/ytterbium oxides and gold nanoparticles

Julia A. Burunkova[1], Ihor Yu. Denisiuk[1], Dmitri I. Zhuk[1], Lajos Daroczi[2], Attila Csik[3], István Csarnovics[*2] and Sándor Kokenyesi[2]

[1]Engineering photonics, ITMO University, Kadetskaja line 3/2, St. Petersburg, Russia
[2]Institute of Physics, University of Debrecen, Bem sq. 18a, Debrecen, 4026, Hungary
[3]Institute for Nuclear Research, Hungarian Academy of Sciences, Bem sq. 18/c, Debrecen, 4026, Hungary

## Abstract

Rare-earth-doped optical materials are important for light sources in optoelectronics, as well as for efficient optical amplification elements and other elements of photonics. On the basis of the previously developed method of anhydrous, low-temperature synthesis of Er/Yb oxides from their chlorides we fabricated proper nanoparticles with defined parameters and used them for the development of optically transparent, luminescent polymer nanocomposite with low optical scattering, suitable for direct, light-induced formation of photonic elements. Introduction of preformed gold nanoparticles in such a nanocomposite was also performed and an enhancement of luminescence due to the influence of plasmon effects was detected.

## Introduction

Rare-earth (RE) ions are widely investigated and applied, e.g., in photonics, as active materials for optical amplification or as biomarkers. Er-doped materials are important for light sources in optoelectronics, especially for the telecommunication window at 1.54 μm, as well as for efficient optical amplification elements. For example, silicon oxides ($SiO_x$) and silicon nitrides ($SiN_x$) doped with Er can be integrated in a metal-oxide-semiconductor structure and used as infrared light sources or amplifiers in telecommunication systems [1]. Planar waveguide amplifiers on the basis of Er-containing nanocomposites are actively investigated as well (see for example [2], where a PMMA-based, $Yb_2O_3/Er_2O_3$-containing polymer nanocomposite waveguide optical amplifier was demonstrated). This material ensures the functioning of the amplifier with an interaction length of about 5 mm, which is sufficient for applications in integrated optics.

It is an important task to create such functional materials with high concentrations of Er for waveguides of integrated optical systems that can realize amplification in a few millimeters long waveguide instead of macroscopic elements. At the same time one of the main disadvantages of nanocomposites is the high level of light scattering, which can be caused by agglomeration of nanoparticles. So the development of low-scattering, transparent nanocomposites is an important challenge nowadays.

The method of thermal decomposition of rare earth salts is usually used for the fabrication of luminescent rare earth oxide (REO) nanoparticles [3]. Unfortunately, the agglomeration of nanoparticles during high temperature annealing (above 900 °C) causes a high level of light scattering in the nanocomposite made from these materials, which makes them unusable for photonics. Other methods also have shortcomings. For example, the synthesis of $Er_2O_3$ nanoparticles through the co-precipitation method with addition of coordinating ligands needs additional co-precipitants, and still high temperature annealing is necessary [4]. This method results in large, i.e., tens of nanometers, REO particles. At the same time it is important to avoid the high level of light scattering in a composite with a rather high concentration of RE ions nanoparticles and to assure a high luminescence output. The simple increase of the concentration of small nanoparticles is not always a good solution to the problem because of the possible decrease of luminescence intensity due to the counterproductive introduction of energy-transfer processes at high concentrations of rare earth ions. For example, as it was shown in [5] for Er-doped silicon materials, a practical limitation to the concentration of RE ions is connected with a small oscillator strength of the $Er^{3+}$ intra-4f transition at 1.54 m ($^4I_{13/2}$ $^4I_{15/2}$)). Loss processes can easily dominate the de-excitation of the $^4I_{13/2}$ level. Cooperative upconversion, which can dominate the de-excitation of the $Er^{3+}$ $^4I_{13/2}$ level even at $Er^{3+}$ concentrations as low as ca. 0.02 atom % [4], results in depopulation by energy exchange between two excited $Er^{3+}$ ions. Also, the concentration quenching in Er-doped silicon is related to energy migration to defect states or OH-groups by resonant interaction between closely spaced $Er^{3+}$ ions [5]. In Er-doped dielectrics, the Er concentration needs to be kept low to avoid concentration quenching and cooperative upconversion. The possible solution can be to increase the excitation efficiency at lower ion concentrations, also in presence of Yb ions. In some works [6-8] the increase of the rare-earth luminescence intensity was detected in the presence of plasmon fields, which were generated around metallic nanoparticles. This effect seems to be applicable for Er-containing polymer nanocomposites as well. At the same time the influence of metallic nanoparticles can be negative, which together with the negative influence of OH radicals on the IR transmissions makes the development of new composites necessary.

The aim of our work was to use the previously developed method of a water-free, low-temperature (below 230 °) synthesis of rare-earth oxide nanoparticles from their chlorides [9] for the fabrication of Er/Yb oxide nanoparticles with defined parameters and without agglomeration, and the fabrication of optically transparent nanocomposites on their basis in selected polymer matrices with tunable luminescent parameters. Introduction of preformed gold nanoparticles and realization of plasmonic effects in such a nanocomposite was also aimed and successfully performed. This can be further used as continuation of our previous work on the investigations of polymer nanocomposites for photonics [9-11].

## Results and Discussion
### Investigations of rare-earth oxide nanoparticles

We propose the reaction scheme in Figure 1 for the synthesis. In this synthesis route $SiO_2$ NPs are necessary components of the synthesis process, because the rare-earth oxide (REO) hydroxides are absorbed just on these particles. Annealing of the obtained REO hydroxides at 230 °C and their transformation into oxides takes place on the surface of $SiO_2$ NPs. Without $SiO_2$ NPs the reaction yield is close to zero, because mainly REO products that are soluble in glycerin, probably glycerolates, are produced (see Figure 1). The advantages of the proposed method in comparison with known ones are the synthesis in a water-free environment and the absence of high-temperature annealing. Low-temperature annealing hinders the coagulation of Er/Yb oxides and no agglomerates are present in the reaction medium as well as in the polymer nanocomposites, as it will be shown later.

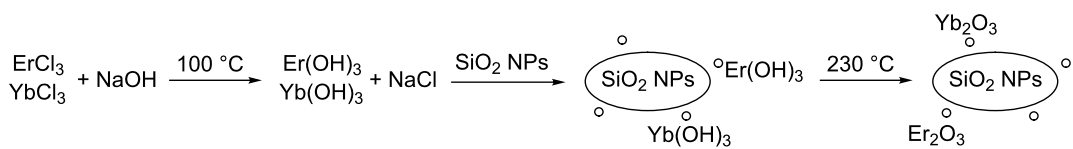

**Figure 1:** The outline of the synthesis process of Er/Yb oxide nanoparticles.

The introduction of $SiO_2$ NPs enables a high yield of REO in the proposed reaction. Synthesis without silica addition results in products that are soluble in glycerin. The presence of silica nanoparticles improves the compatibility of the obtained REO particles with the polymer matrix when fabricating nanocomposites. At the same time the luminescent properties are not impaired, which it will be shown later. The synthesis was developed for obtaining a mixture of erbium and ytterbium oxides. The combination of the two elements is requried, in order to be able to pump at 980 nm and to obtain luminescence at 1500 nm. This is because the direct absorption of erbium ions at 980 nm has a low efficiency, but for ytterbium ions it is highly efficient and the ytterbium ions can transfer excitation energy to the erbium ions.

Energy dispersive X-ray analysis (EDX) in a scanning electron microscope (SEM) was used for determining the composition of the obtained oxide powders. The results show an almost total transformation of Er and Yb chlorides into oxides during the synthesis, only traces of chlorides were detected (Table 1).

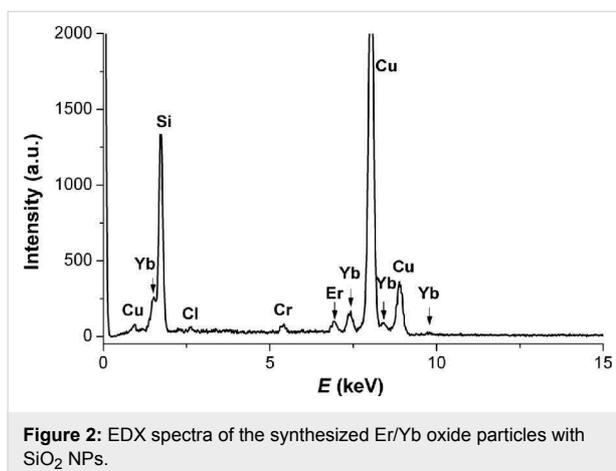

**Figure 2:** EDX spectra of the synthesized Er/Yb oxide particles with $SiO_2$ NPs.

**Table 1:** Quantitative data of composition analyses.

| element | initial mixture of components | | synthesized powder | |
|---|---|---|---|---|
| | wt % | atom % | wt % | atom % |
| Er | 6.27 | 0.93 | 5.84 | 0.83 |
| Yb | 13.30 | 1.90 | 12.45 | 1.72 |
| Si | 29.14 | 25.72 | 37.68 | 31.99 |
| O | 41.85 | 64.84 | 43.82 | 65.32 |
| Cl | 9.44 | 6.60 | 0.21 | 0.14 |

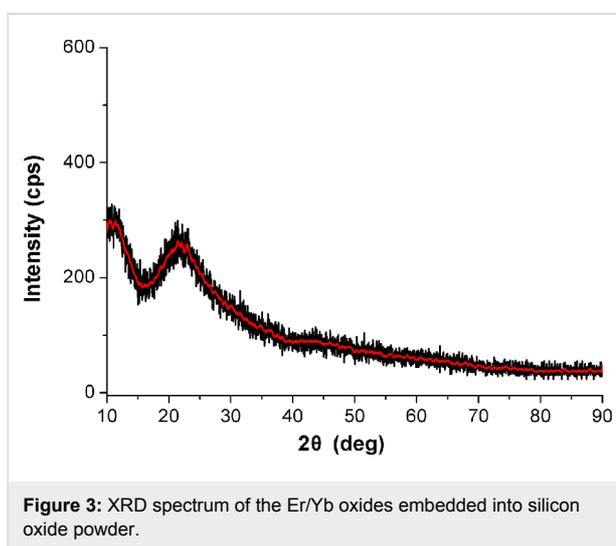

**Figure 3:** XRD spectrum of the Er/Yb oxides embedded into silicon oxide powder.

This composition was also corroborated in further transmission electron microscope (TEM) analyses of the nanocomposites (see below in Figure 6), although oxygen is not visible in the EDS spectra due to the parameters of TEM, which are mentioned in the Experimental section (Figure 2).

Summarizing the results it can be noted, that the water-free low-temperature synthesis yielded nanoparticles of Er/Yb oxides with a mass ratio of 1:2, embedded in silicon oxide powder, the mass of which is twice as high as that of the REO. This corresponds well to the desired composition. No crystalline phases were detected in this powder (see Figure 3), however, it is possible that the sample contains crystals with sizes of 3–4 nm or below. This is the limit for X-ray diffraction (XRD), and in this case the sample will look like an amorphous phase. Regardless of this, the powder can be used it for the preparation of polymer nanocomposites.

One of the main issues of this work was to obtain luminescence, and to measure the spectra of luminescence excited at 488 nm for the initial mixture of Er/Yb chlorides, as well as of the powder of Er/Yb oxides, obtained with and without $SiO_2$ NPs. These spectra are presented in Figure 4.

It can be seen from the data in Figure 4, that the obtained powder mixture, as well as its dispersion in glycerin exhibit much stronger luminescence than the initial mixture of chlorides. The measurement with an optical microscope had shown that the dispersion of the luminescent particles in glycerin is uniform. That supports that no agglomeration of Er/Yb oxide particles occurred during the water-free low-temperature synthesis. It is also clear, that the maximum of the luminescence of Er/Yb oxide obtained without or with $SiO_2$ NPs is located near 556 nm (Figure 4b). This corresponds to the transition $^4S_{3/2}$ $^4I_{15/2}$ and produces a rather broad maximum, usually observed in luminescence spectra of Er in a transparent matrix. It appears due to the superposition of separate Er-ion lines in the ed and described in literature, were not especially targeted here, although IR luminescence signals were detected in polymer nanocomposites (see below in Figure 7).

## Investigations of polymer nanocomposites with Er and Yb oxides and gold nanoparticles

On the basis of the described synthesis, optically transparent colorless or deep-reddish monomer nanocomposites were obtained (Table 2). After the UV-curing of the monomer layer placed between substrate and cover we obtained colorless (without AuNPs) or red (with AuNPs) optically transparent homogeneous films with thicknesses in the range of 25–150 μm. A composition consisting of urethane monomers urethane dimethacrylate (UDMA) and isodecyl acrylate (IDA) with a saturated hydrocarbon side chain was selected for the polymer matrix. The application of monomers with nitrogen atoms in the structure and the presence of the saturated hydrocarbon chain in the monomer, as well as the introduction of $SiO_2$ NPs, allowed us to obtain homogeneous mixtures with rare-earth oxide and gold nanoparticles (AuNPs). Something that could not be successfully realized with matrix materials used before [9].

Examples of absorbance spectra of these composites are presented in Figure 5. Slight absorption bands of erbium and ytterbium oxides at 530 and 980 nm are visible in spectrum 1. A strong absorption band at 530 nm appears after the introduction of AuNPs to the composite. It is evidently related to the resonant optical absorbance of localized plasmons in the AuNPs of the given size.

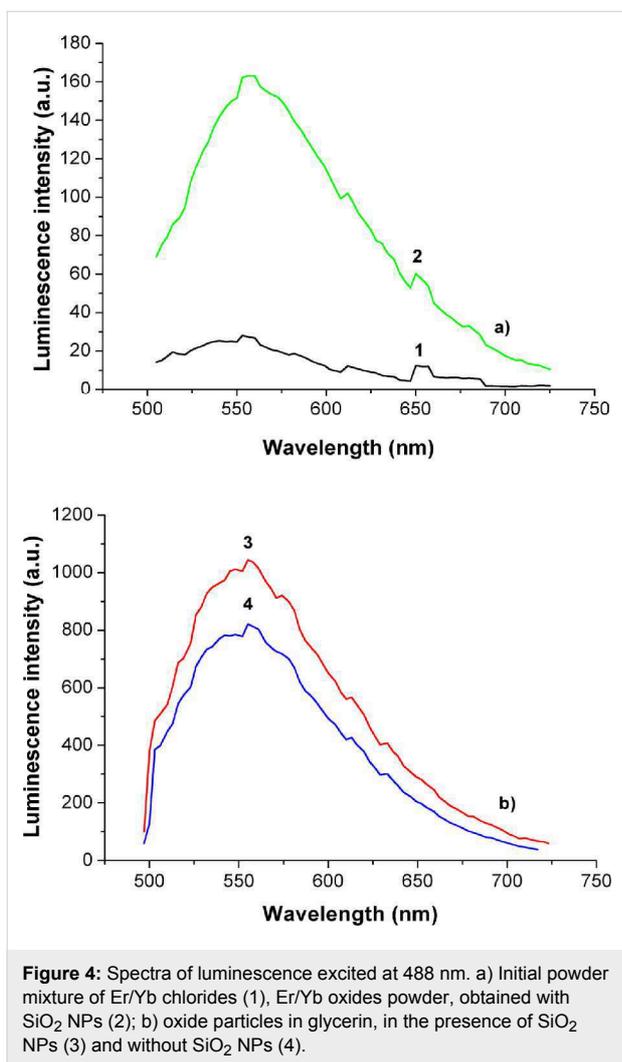

**Figure 4:** Spectra of luminescence excited at 488 nm. a) Initial powder mixture of Er/Yb chlorides (1), Er/Yb oxides powder, obtained with $SiO_2$ NPs (2); b) oxide particles in glycerin, in the presence of $SiO_2$ NPs (3) and without $SiO_2$ NPs (4).

mixture of non-uniform particles, associated with silica, which in turn should be introduced to the synthesis. Otherwise the reaction yield of the synthesis without $SiO_2$ NPs is smaller by an order of magnitude. The excitation, luminescence and upconversion effects in the near-IR spectral region, usually investigat-

TEM investigations confirm (Figure 6) the rather homogeneous distribution of nanoparticles in the polymer matrix. According to our assumptions and previous investigations [12] the gold nanoparticles should be located around the silicon oxide nanoparticles, which prevents their agglomeration up to 0.15 wt % of content and the plasmon resonance absorbance spectra are observable in the polymer nanocomposite.

**Table 2:** Composition of the prepared composites.

| sample no. | monomers | $SiO_2$ NPs wt % | Au NPs wt % | Er/Yb oxides wt % |
|---|---|---|---|---|
| 14a | UDMA/IDA = 3:7 | 10 | — | — |
| 14Au | UDMA/IDA = 3:7 | 10 | 0.15 | — |
| 18Au | UDMA/IDA = 3:7 | 10 | 0.30 | — |
| 50a | UDMA/IDA = 3:7 | 10 | — | 0.17 |
| 50Au | UDMA/IDA = 3:7 | 10 | 0.15 | 0.17 |
| 52Au | UDMA/IDA = 3:7 | 10 | 0.15 | 0.50 |
| 54 | UDMA/IDA = 3:7 | 10 | — | 1.7 |
| 54Au | UDMA/IDA = 3:7 | 10 | 0.15 | 1.7 |

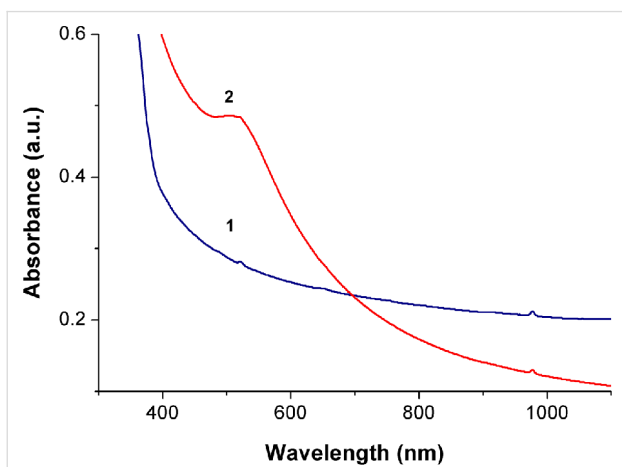

**Figure 5:** Optical absorbance spectra of composites 54a (1) and 54Au (2).

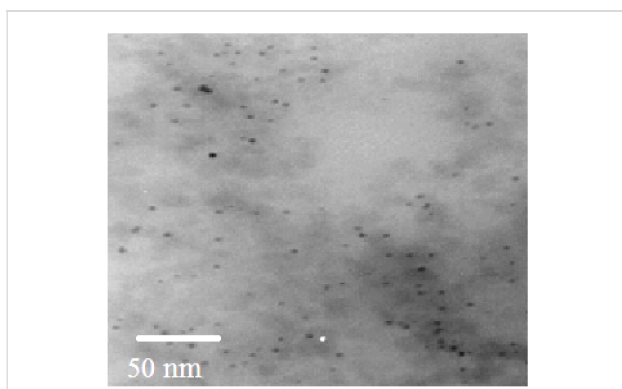

**Figure 6:** TEM pictures of the nanocomposite film REO-AuNPs-SiO$_2$NPs-UDMA/IDA (52Au sample).

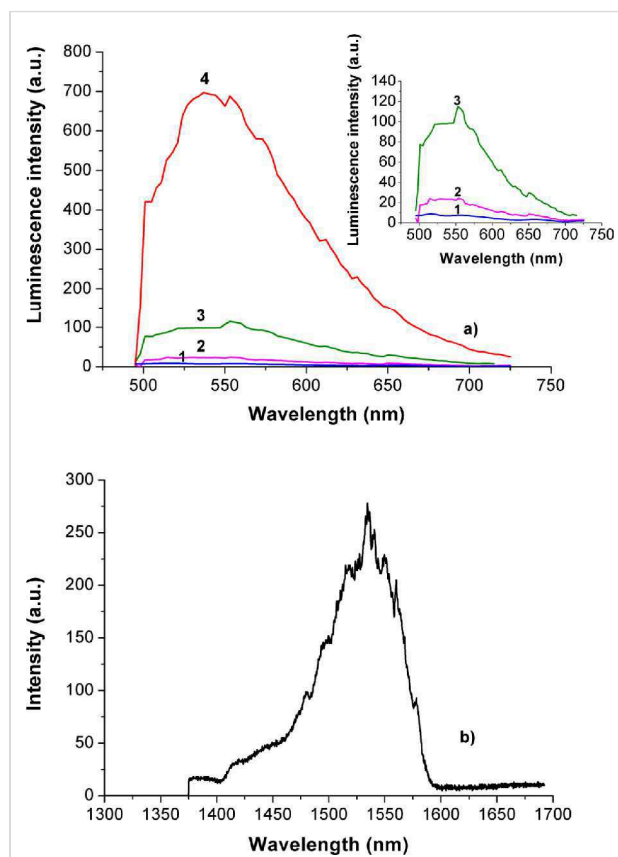

**Figure 7:** Luminescence spectra of the polymer nanocomposites films. a) Excitation at 488 nm: 1 - UDMA/IDA + 10 wt % SiO$_2$ NPs (14a); 2 - UDMA/IDA + 10 wt % SiO$_2$ NPs + 0.15 wt % AuNPs (14Au); 3 - UDMA/IDA + 10 wt % SiO$_2$ NPs + 0.17 wt % Er/Yb oxide NPs (50a) 4 - UDMA/IDA + 10 wt % SiO$_2$ NPs + 0.17 wt % Er/Yb oxide NPs + 0.15 wt % AuNPs (50Au) b) Excitation at 980 nm: UDMA/IDA + 10 wt % SiO$_2$ NPs + 1.7 wt % Er/Yb oxide NPs (54a).

The measurement with a laser scanning microscope (pump wavelength of 488 nm) has shown that the Er luminescence centers are homogeneously distributed and located in the regions of amplified optical fields near the excited AuNPs, so the amplification of luminescence is possible. Luminescence spectra of REO nanocomposites with and without AuNPs are presented in Figure 7. The broadening of the luminescent band may be connected with peculiarities of the nanocomposite structure, where REO are formed at the surface of amorphous SiO$_2$ NPs with random distribution, which results in slightly different locations of emitting centers in the local fields. The Stark splitting and structure of spectra result from the superposition of different lines of different emitting centers.

If the erbium luminescence centers are homogeneously distributed and located in random regions of the amplified optical field near the excited AuNPs, the amplification of luminescence is possible, like it was observed and modeled in [6] for nanocomposites with silver nanoparticles and different distances between ions and silver nanoparticles in the range between 10 and 300 nm. Indeed, as it follows from our results (Figure 7) enhancement of the luminescence output was observed in nanocomposites with AuNPs. This effect can be even larger, since the luminescence excitation wavelength was only at the short wavelength side of the plasmon resonance spectrum. But at the same time the emitted light is just in the optimum range for such excitation, so certain self-enhancement effects take place in our nanocomposites with AuNPs. Direct optical microscopy investigations supported the uniform distribution of luminescence from the surface of the excited layer, which correlates with TEM data in Figure 6.

Obviously, since the luminescent Er/Yb oxide nanoparticles are located near the surface of silica nanoparticles, which fill the volume of the composite rather uniformly because of our synthesis technology, and the gold nanoparticles are also attached there, the plasmon field of excited AuNPs can enhance the quantum yield of luminescence. Of course, they should be sepa-rated by 5–10 nm, to prevent tunneling of the excited electrons from the luminescence center to the surface of the gold nanoparticles. Such a separation of gold and oxide nanoparticles is realized by the separated absorbance of both types of nanoparticles on the SiO$_2$ NPs with a radius of 7 nm. It is also not excluded that the distance between some SiO$_2$-Au-REO NP complexes randomly distributed in polymer matrix falls within tens of nanometers, where the influence of plasmon fields on RE ion excitation is efficient.

Based on the earlier data on direct surface patterning in polymer nanocomposites [10-12] the in situ formation of two-dimensional photonic structures with luminescent properties (spectra similar to Figure 7) was performed by the method of four-beam holography (Figure 8). Such structures can be used for integrated sensors and other photonic elements.

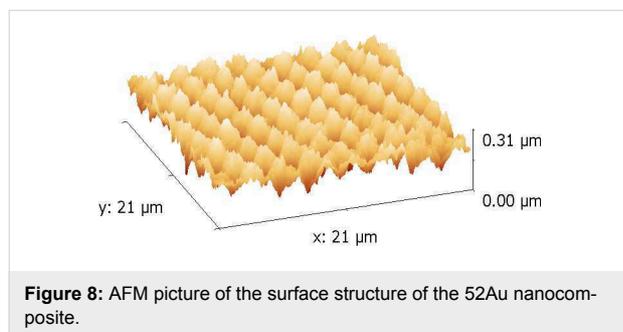

**Figure 8:** AFM picture of the surface structure of the 52Au nanocomposite.

## Conclusion

Er/Yb oxide nanoparticles were fabricated by a water-free low-temperature synthesis ($T$ 230 °C) in the presence of SiO$_2$ nanoparticles. This technology route allows one to avoid high-temperature annealing. These conditions ensure the absence of agglomeration of Er/Yb oxide particles not only in the reaction medium (glycerin), but also in the polymer nanocomposites. The use of SiO$_2$ nanoparticles increases the reaction yield of this technology process.

Polymer nanocomposites containing Er/Yb oxide nanoparticles as well as SiO$_2$ nanoparticles and Au nanoparticles were created. SiO$_2$ nanoparticles compatible with the selected polymer matrix can be used to ensure a uniform distribution of Au and Er/Yb oxide nanoparticles within the matrix polymer. Absorbance bands of Er oxide and AuNPs plasmon resonance absorbance are present in the spectra of nanocomposites.

Er luminescence in nanocomposites is enhanced by almost one order of magnitude in the presence of plasmon fields of excited AuNPs, and it is important to note, the effect of self-enhancement of luminescence in such a gold-containing nanocomposite is possible.

## Experimental
### Materials and investigation methods

The following materials and chemicals were used in this work: diurethane dimethacrylate, mixture of isomers (436909 ALDRICH, UDMA); isodecyl acrylate (408956 ALDRICH, IDA); initiator 2,2-dimethoxy-2-phenylacetophenone (19611-8 Aldrich, In2); dodecanethiol-functionalized gold nanoparticles, 5 nm (Nanoprobes, no. 3014, AuNPs); erbium chloride (Aldrich, 449792, ErCl$_3$); ytterbium chloride (Aldrich, 439614, YbCl$_3$); SiO$_2$ nanoparticles, 7 nm in diameter (Aldrich no. 066K0110, SiO$_2$ NPs); glycerol, sodium hydroxide; 1-butanol.

The resulting nanoparticles and nanocomposites were characterized by TEM (JEOL-2000FXII equipped with Oxford Link Be window X-Ray spectrometer). Despite the good topographical imaging of the structure this equipment cannot detect oxygen, so additional EDS investigations were performed by using a Bruker spectrometer in a Hitachi S-4300 SEM.

Optical spectra were measured with a Shimadzu 1800 UV–vis spectrometer. The X-ray diffraction (XRD) measurements were carried out by using a diffractometer using Siemens Cu-anode X-ray tube and a horizontal goniometer equipped with a graphite monochromator. The high angle spectra were measured between 10–90° to study the presence of crystalline phase in the samples.

Luminescence measurements of the obtained Er/Yb oxide nanoparticles as well as of polymer nanocomposites on their basis were performed by the scanning laser microscope Zeiss LSM-710 with pumping light at 488 nm and 3 nm spectral steps. Surface reliefs of the structures, optically recorded in nanocomposites were mapped by using AFM (Veeco diCaliber).

### Synthesis of Er/Yb oxide nanoparticles

We have further developed two routes [9] for the preparation of rare earth oxides (Er and Yb) (REO), using silicon oxide nanoparticles (SiO$_2$ NPs) during the synthesis or not. The process was the following: The mixture of erbium and ytterbium chlorides in proportion 1:2 was dissolved at 100 °C in glycerin (RE salts/glycerin = 1 g/60 mL). Further, 5% solution of NaOH in glycerin was added under intensive mixing. The amount of alkaline was calculated as 0.43 g alkaline for 1 g mixture of RE salts. SiO$_2$ NPs with total mass two times larger as of the RE salts were added immediately after adding alkaline.

In the case of the second preparation route for rare earth oxides SiO$_2$ NPs were not added. In both cases the solution was heated up to 140 °C and kept at this temperature for 1 h, as well as at 180 °C for 4 h. After cooling, the solution was repeatedly washed in butanol to reach pH 7 and dried at 50 °C. The ob-tained powder was dissolved in glycerin and heat treated at 230 °C for 1 h. After cooling, the sediment was repeatedly washed in butanol and dried at 50 °C. The result was a white powder.

## Synthesis of nanocomposites, containing synthetized REO and AuNPs

UDMA and IDA monomers were intensively mixed for 3 h and after this toluene was added. The powders of Er and Yb oxides and $SiO_2$ NPs were introduced directly to the toluene solution of monomers, the mass concentration of RE oxides was between 0.17 and 5 wt %. 10 wt % of $SiO_2$ NPs were added and followed by 10 h intermixing. After this step AuNPs in toluene solution with a concentration between 0.08 and 0.15 wt % was added and the solution was mixed for 10 h. Introduction of $SiO_2$ NPs enhances compatibility of AuNPs and REO with the organic matrix.

The next step was adding 2,2-dimethoxy-2-phenylacetophenone as photo-initiator at a concentration of 0.5 wt %. Thus, optically transparent colorless or red monomer nanocomposites were prepared (Table 1). After spreading on a substrate and UV curing colorless (without AuNPs) or red (with AuNPs) transparent uniform films were fabricated.

## Acknowledgements

This work was supported by the International laboratory "Nonlinear optical molecular crystals and microlasers" project 713551 ITMO University 5/100, by Ministry of Education of Russian Federation project no. 3.432.2014/K, as well as by the TAMOP 4.2.2.A-11/1/KONV-2012-0036 project, which is co-financed by the European Union and European Social Fund.